\documentstyle[12pt,epsfig]{article}
\begin{document}
\begin{titlepage}
\thispagestyle{empty}

\begin{flushright}
 UCLA/98/TEP/12\\
March 1998\\
\end{flushright}

\vskip 2.cm
\begin{center}
{\Large\bf SU(2) Poisson-Lie T duality \renewcommand{\thefootnote}{\fnsymbol{footnote}}\footnote{Work supported in part by CONICIT fellowship and by DOE grant DE-FG03-91ER40662}}
\vskip 2.cm
{\large  M.A. Lled\'o$^{a,b}$ and V.S. Varadarajan$^c$}
\end{center}

{\it $^{a}$Centro de F\'{\i}sica. Instituto Venezolano de Investigaciones Cient\'{\i}ficas (IVIC). Apdo 21827 Caracas 1020-A. Venezuela.}

{\it $^b$Physics Department and  $^c$Mathematics Department, University of California, Los Angeles. 405 Hilgard Av. Los Angeles, CA 90095-1547. USA.} 

\begin{abstract}
Poisson-Lie target space duality is a framework where duality transformations are properly defined. In this letter we investigate the dual pair of $\sigma$-models defined by the double SO(3,1) in the Iwasawa decomposition.\end{abstract}
\end{titlepage}

\section{Introduction}

Target space duality was known as a property of two dimensional sigma models with isometries. (For a review, see Ref.\cite{aal} and references inside). Abelian duality is possible when the group of isometries of the target manifold is abelian . In this case \cite{cfgd} \cite{rv}, after gauging the isometry with a flat connection and integrating out  the gauge fields one obtains an equivalent theory defined in a target space with different background fields. The dual theory has also abelian isometries and when the duality transformation is performed again, the original model is recovered. This scheme was generalized \cite{oq} to  consider non abelian groups of isometries. The dual model is obtained in the same way, but it may not have isometries anymore, so it is not possible to perform the duality transformation again to obtain the original model. It was clear then that duality should be understood in a more general framework where isometries will not play a fundamental role. In fact, duality transformations can be seen as canonical, (perhaps non local)  changes of variables \cite{cz}, making manifest the unnecessary role of isometries.

A step forward in order to clarify this question was given in Ref.\cite{ks1}. The condition of symmetry was substituted by a weaker one. The model is not required to be symmetric, but there is still an action of a group $G$ on the manifold and the Noetherian currents associated to this action are required to be integrable, that is, they are the components of a flat connection with a certain structure group $G^*$ with the same dimension than $G$. The duality transformation is then well defined and reflexive for the class of all models satisfying this property. The groups $G$ and $G^*$ can not be chosen randomly; instead, they have to satisfy a compability condition, their Lie algebras are dual to each other  in the sense that they define a bialgebra structure. The infinitesimal notion of bialgebra structure corresponds, when exponentiating to the group, to a Poisson-Lie structure. A Poisson-Lie structure is the classical limit of a quantum group, so we can see that duality transformations give rise to a rich structure that ultimately may be related to a quantum group. 

Few models \cite{akt,k} have been explicitely constructed that are not examples of the original semi-abelian duality (that is, when the model has isometries so the dual group is abelian). In this letter we want to construct a new model. The group $G$ is SU(2). The action on the target manifold is transitive and free, so the group itself can be identified with the target manifold. The bialgebra structure is the standard one; then Drinfeld's double is just SO(3,1) $\approx$ Sl(2,c)$^r$. The dual group is solv(Sl(2,c)$^r$), given by the Iwasawa decomposition of Sl(2,c)$^r$.

\section{Poisson-Lie T duality}

In this section we summarize the basic facts about Poisson-Lie T-duality \cite{ks1, akt}
We consider a  two-dimensional $\sigma$-model on a target manifold $M$ described by the action
\begin{equation}
S=\int dzd\bar z(g_{ij}+b_{ij})\partial x^i\bar \partial x^j=\int dzd\bar zE_{ij}\partial x^i\bar \partial x^j
\end{equation}
where $g_{ij}$ is the metric on $M$ and $b_{ij}$ is an antisymmetric tensor. The null coordinates $z$ and $\bar z$ are given by 
\begin{equation} 
z={1\over\sqrt 2}(\xi^0+\xi^1),\qquad \bar z={1\over\sqrt 2}(\xi^0-\xi^1)
\end{equation}
We assume that a Lie group $G$ acts transitively on $M$ (not necessarily being an isometry) and let $v_a(x), a=1,\dots n$ the vector fields generators of the action of $G$ on $M$. Under the infinitesimal transformations
\begin{equation} \delta x^i=\epsilon(x)^av_a^i(x) \end{equation}
the action changes
\begin{equation}\delta S=\int  dzd\bar z\bigr\{\epsilon^a{\mathcal L}_{v_a}E_{ij}\partial x^i\bar \partial x^j\bigl\}-\int\epsilon^a\wedge dJ_a\end{equation}
where $J_a$ are the Noetherian currents associated to the action of $G$, and are given by
\begin{equation}
J_a=E_{ij}v^i_a\bar\partial x^jd\bar z-E_{ij}v^j_a\partial x^idz
\end{equation}
On the extremes of $S$, $\delta S=0$,  one has
\begin{equation}
dJ_a={\mathcal L}_{v_a}E_{ij}\partial x^i\bar \partial x^j
\label{bi}\end{equation} 
and if ${\mathcal L}_{v_a}E_{ij}=0$ the currents are conserved. But we can relax the condition of symmetry and require instead that the currents are the components of a flat connection with certain structure group $G^*$
\begin{equation}dJ_a={1\over 2} \gamma_a^{bc}J_b \wedge J_c
\end{equation}
condition that is solved if $E_{ij}$ satisfies
\begin{equation}
{\mathcal L}_{v_a}E_{ij}=\gamma_a^{bc}v_b^kv_c^lE_{kj}E_{il}\label{sym}
\end{equation}
Note that the symmetric case is only a particular case when the group $G^*$ is abelian. Surprisingly enough, the  duality transformation is a well defined equivalence  in the class of models satisfying this condition.

Notice that there is a compatibility condition for Eq.(\ref{sym}) when applying twice the Lie derivative. If $c_{ab}^c$ are the structure constants of the group $G$, then the relation between both sets of structure constants
\begin{equation}
c_{rs}^t\gamma_t^{ab}=c_{rp}^a\gamma_s^{pb}+c_{rq}^b\gamma_s^{aq}-c_{sp}^a\gamma_r^{pb}-c_{sq}^b\gamma_r^{aq}
\end{equation}
should be satisfied. As we will see in next section, this defines a Lie bialgebra structure on $G$. It is shown in \cite{ks1} that there exists a dual (equivalent) $\sigma$-model defined by a matrix $\tilde E_{ij}$ where the role of groups $G$ and $G^*$ is interchanged, that is, there is an action of $G^*$ on the target manifold of the dual model such that if $\tilde v_a$ are the generators of this action then
\begin{equation}
{\mathcal L}_{v_a}\tilde E_{ij}=\gamma_a^{bc} \tilde v_b^k\tilde v_c^l\tilde E_{kj}\tilde E_{il}
\end{equation}

\section{Manin triples and Iwasawa decomposition \label{Iwa}}

We briefly recall here the definition of a Lie bialgebra in terms of Manin triples \cite{cp}. In this paper we shall consider only finite dimensional Lie algebras.

A Manin triple is a triple of Lie algebras $(\mathcal{P}, \mathcal{P}_+, \mathcal{P}_-)$ and a symmetric, non degenerate bilinear form $B$ on $\mathcal{P}$  such that

\noindent 1. $ {\mathcal P}_+$ and $ {\mathcal P}_-$ are Lie subalgebras of $ \mathcal{P}$,

\noindent 2. $  {\mathcal P}={\mathcal P}_+\oplus  {\mathcal P}_-$ as vector spaces, 

\noindent 3. $ {\mathcal P}_+$ and ${\mathcal P}_-$ are isotropic for $B$.

The bilinear form $B$ determines an isomorphism between ${\mathcal P}_+^*$ (the dual vector space of ${\mathcal P}_+$)  and ${\mathcal P}_-$. On 
${\mathcal P}\approx {\mathcal P}_+\oplus  {\mathcal P}_+^*$, $B$ is given by
\begin{equation}
B(X,Y)=0,\quad B(\zeta,\eta)=0,\quad B(\zeta,X)= \zeta (X),\qquad 
\zeta,\eta\in{\mathcal P}_+^*;X,Y \in{\mathcal P}_+
\end{equation}
The isomorphism  induces a Lie algebra structure on $\mathcal{P}_+^*$ (the dual Lie algebra of ${\mathcal P}$, $[\;,\;]_{{\mathcal P}_+^*}$. This is equivalent to give a cocommutator on ${\mathcal P}_+$
\begin{equation}
\delta_{{\mathcal P}_+} :{\mathcal P}_+\to{\mathcal P}_+\otimes{\mathcal P}_+
\end{equation}
such that
\begin{equation}
[\zeta,\eta]_{{\mathcal P}_+^*}(X)=\delta_{{\mathcal P}_+}(X)(\zeta,\eta),\qquad \zeta,\eta\in{\mathcal P}_+^*;X\in{\mathcal P}_+
\end{equation}
The Jacobi identity for the  Lie bracket on ${\mathcal P}$ (that leaves invariant the bilinear form $B$) is equivalent to the fact that $\delta_{{\mathcal P}_+}$ is a cocycle of the Lie algebra ${{\mathcal P}_+}$ with values in ${{\mathcal P}_+}\otimes{{\mathcal P}_+}$. So we arrive (for finite dimensional Lie algebras) to two equivalent definitions of Lie bialgebra, in terms of Manin triples or in terms of the cocommutator.

Choosing a basis $\{X_r, r=1,\dots n\}$ in ${\mathcal P}_+$ and the corresponding dual basis in ${\mathcal P}_+^*$  $\{\zeta^r, r=1,\dots n\}$ we have
\begin{equation}
[X_r,X_s] = c_{rs}^tX_t, \quad [\zeta^r,\zeta^s] = \gamma_t^{rs}\zeta^t ,\quad[\zeta^r,X_s] = c_{st}^r\zeta^t-\gamma^{st}_rX_t
\end{equation}
and the cocycle condition is simply
\begin{equation}
c_{rs}^t\gamma_t^{ab}=c_{rp}^a\gamma_s^{pb}+c_{rq}^b\gamma_s^{aq}-c_{sp}^a\gamma_r^{pb}-c_{sq}^b\gamma_r^{aq}
\end{equation}

The Lie algebra ${\mathcal P}$ has a bialgebra structure itself given by
\begin{equation}
\delta_{{\mathcal P}}(X_s)=\gamma_s^{rt}X_r\wedge X_t,\qquad\delta_{{\mathcal P}}(\zeta^s)=c^s_{rt}\zeta^r\wedge \zeta^t
\end{equation}
The inclusions ${\mathcal P}_+\subset{\mathcal P}$ and $({\mathcal P}_-)^{\hbox {\footnotesize op}}\subset{\mathcal P}$ are homomorphisms of Lie bialgebras. With this bialgebra structure, ${\mathcal P}$ is the double of ${\mathcal P}_+$, and the dual of ${\mathcal P}$ is isomorphic to the direct sum as Lie algebras ${\mathcal P}={\mathcal P}_+\oplus{\mathcal P}_-$

We want now to describe the structure of a double for the Lie algebra  ${\mathcal P}=$sl(2,$c$)$^r$.  Let $\{H,X^+,X^-\}$ be  the standard c-basis for sl(2,c). The commutation relations are
\begin{equation}
[X^+,X^-]=H, \qquad [H,X^+]=2X^+,\qquad [H,X^-]=-2X^-
\end{equation}

The compact real form of sl(2,$c$) is ${\mathcal K}=$sU(2).  ${\mathcal K}$ is the r-linear span of \begin{equation}
\bigl\{G=iH,\quad E=X^+-X^-,\quad F=i(X^++X^-)\bigr\}
\end{equation}
with commutation rules
\begin{equation}
[E,F]=2G,\qquad [G,F]=-2E,\qquad [G,E]=2F
\end{equation}
In terms of  the Pauli matrices
\begin{equation}
\sigma_1=\pmatrix{ 0 & 1 \cr 1 & 0},\qquad\sigma_2=\pmatrix{ 0 & -i \cr i & 0}\qquad \sigma_3=\pmatrix{ 1 & 0 \cr 0 & -1}\label{pauli}
\end{equation}
we can write
\begin{equation}
\bigl\{G=i\sigma_3,\quad  E=i\sigma_2,\quad F=i\sigma_1\bigr\}
\end{equation}

We will describe the double structure in terms of a Manin triple. In order to do that, we consider the Cartan  decomposition \cite{h} of sl(2,$c$)$^r$,
\begin{equation} 
{\mathcal P}={\mathcal K}+i{\mathcal K}
\end{equation}
${\mathcal A}=\hbox{span}\{H\}$ is a maximal abelian subspace of $i{\mathcal K}$. With respect to ${\mathcal A}$ there are two restricted roots, $\lambda_+(H)=2$ and $\lambda_-(H)=-2$ with root subspaces ${\mathcal E}_+=\hbox{span}\{X^+, iX^+\}$ and ${\mathcal E}_-=\hbox{span}\{X^-, iX^-\}$. The Iwasawa decomposition \cite{h} of sl(2,$c$)$^r$ is given by
\begin{equation}
{\mathcal P}={\mathcal K}+\mathcal{A}+{\mathcal E}_+
\end{equation}
The vector space ${\mathcal B}={\mathcal A}+{\mathcal E}_+$ is  a solvable Lie subalgebra of ${\mathcal P}$ and the triple $({\mathcal P}, {\mathcal K}, {\mathcal B})$ is a Manin triple with the bilinear form $B$ given by (minus) the imaginary part of the standard inner product on sl(2,$c$)
\begin{equation}
(X_1,X_2)_{\mathcal{P}}=-\hbox{Im}(X_1,X_2)
\end{equation}
with
\begin{equation} (H,H)=2,\qquad (H,X^\pm)=0,\qquad (X^\pm,X^\pm)=0,\qquad (X^+,X^-)=1
\end{equation}
Identifying ${\mathcal B}$ with ${\mathcal K}^*$ by means of the bilinear form $B$, the dual basis is given by 
\begin{equation}
\bigl\{\tilde G=-{1\over 2}H,\quad  \tilde E=iX^+,\quad \tilde F=-X^+\bigr\}
\end{equation}
and the commutation relations in ${\mathcal B}$ are
\begin{equation}
[\tilde E, \tilde F]=0,\qquad [\tilde E,\tilde G]=\tilde E,\qquad [\tilde F,\tilde G]=\tilde F
\end{equation}

The cocommutator in  $\mathcal{K}$ is then
\begin{equation}
\delta_{{\mathcal K}}(G)=0\qquad \delta_{{\mathcal K}}(E)=E\wedge G,\qquad \delta_{{\mathcal K}}(F)=F\wedge G
\end{equation}
The cocycle $\delta_{{\mathcal K}}$ is a coboundary since there is a $r$-matrix $r\in {\mathcal K}\otimes {\mathcal K}$ 
\begin{equation} r= {1\over 2}E\wedge F\end{equation}
such that
\begin{equation} \delta_{{\mathcal K}}{X}=(\hbox{ad}_X\otimes 1+ 1\otimes\hbox{ad}_X)r\end{equation}

\section{SU(2) $\sigma$-model}

We want to consider now a $\sigma$-model whose target manifold is the group $G$ itself. Eq.(\ref{sym}) can be explicitely solved in that case \cite{ks1, ks2} for the matrix $E_{ij}$.
 
If $G$ is connected and simply connected, any cocycle on $\mathcal{G}$ (the Lie algebra of $G$) with values in $\mathcal{G}\otimes\mathcal{G}$  can be integrated to a cocycle in $G$ with values in  $\mathcal{G}\otimes\mathcal{G}$. Given the cocommutator $\delta_{\mathcal{K}}$ on $\mathcal{G}$, we denote by $\Pi^R:G\to \mathcal{G}\otimes\mathcal{G}$ the corresponding cocycle in $G$. It satisfies the relation
\begin{equation}\Pi^R_{gg'}=\hbox{Ad}_g\otimes \hbox{Ad}_g\Pi^R_{g'}+\Pi^R_{g}\label{pb}
\end{equation}
The bivector $\Pi:G\to TG\times  TG$ translated from $\Pi^R$ by the right action of $G$, $\Pi_{g}=(R_g^T\otimes R_g^T)\Pi^R_{g}$ is a Poisson structure in $G$ compatible with the group law (the compatibility is given by Eq.(\ref{pb})).

Let us assume that the matrix $E_{ij}$ has an inverse $F^{ij}$. Then we can multiply Eq.(\ref{sym}) by $F$ twice and using the fact that the Lie derivative commutes with contractions we obtain
\begin{equation}
{\mathcal L}_{v^L_a}F^{ij}=-\gamma_a^{bc}v^{Li}_bv^{Lj}_c
\label{symf}\end{equation}
($v^L_a$ are the left invariant vector fields, generators of the right action of $G$ on $G$). The matrix $F$ is a bivector in $TG\times  TG$, so it can be written as 
\begin{equation}
F=F^{ab}_gv^R_a\otimes v^R_b=F^{ab}_gR_g^TX_a\otimes R_g^TX_b
\end{equation}
where $v^R_a$ are the right invariant vector fields and $X_a$ are a basis in $\mathcal{G}$. It is obvious that to any solution of Eq.(\ref{symf}) we can add a constant matrix $R^{ab}v^R_a\otimes v^R_b$. It is also easy to prove that $F^{ab}_g=-\Pi^{Rab}_g$ is a solution to Eq.(\ref{symf}).

We want to compute the Poisson bivector in SU(2). An element $x$ of SU(2) is given by a matrix
\begin{equation}
x=x_0I-i\vec{x}\cdot\vec{\sigma}
\end{equation}
where $\vec{x}=(x_1,x_2,x_3)$,  $\vec{\sigma}=(\sigma_1,\sigma_2,\sigma_3)$ are the Pauli matrices of Eq.({\ref{pauli}) and $I$ is the identity matrix. The determinant condition is equivalent to
\begin{equation}
x_0^2 +x_1^2+x_2^2+x_3^2=1\end{equation}
The group acts linearly on R$^4$. The left and right actions are
\begin{equation}
L_x=\pmatrix{ x_0 & -x_1  & -x_2 & -x_3\cr x_1 &x_0 & -x_3 & x_2 \cr x_2& x_3 &x_0 & -x_1 \cr x_3 & -x_2 & x_1 &x_0} \qquad R_x=\pmatrix{ x_0 & -x_1  & -x_2 & -x_3\cr x_1 &x_0  & x_3 & -x_2 \cr x_2& -x_3 &x_0 & x_1 \cr x_3 & x_2 & -x_1 &x_0}
\end{equation}

As we remarked in section \ref{Iwa} the coalgebra structure on sU(2) is coboundary and the Poisson structure is given by the Sklyanin bracket

\begin{equation}
\Pi(x)=(L_x^T\otimes L_x^T-R_x^T\otimes R_x^T)r
\end{equation}
with
\begin{equation}
r={a\over 2} E\wedge F\label{pbi}\end{equation}
where $a$ is any real number. Explicitely it becomes

\begin{equation}
\Pi(x)=\Pi^{ij}(x)\partial_i\otimes\partial_j
\end{equation}

\begin{equation}
\Pi^{ij}(x)=a\pmatrix{ 0 & -x_1x_3  & -x_2x_3 & x_1^2+x_2^2\cr x_1x_3 &0  & 0  &-x_0 x_1 \cr x_2x_3& 0 &0 & -x_0x_2 \cr -(x_1^2+x_2^2)  & x_0x_1 & x_0x_2 &0}\end{equation}

Any Poisson-Lie structure is always degenerate, so the matrix $\Pi^{ij}(x)$ cannot be inverted. We need to add a matrix $R^{ij}$. We choose the inverse of the natural invariant metric on SU(2). Our model is then given by 
\begin{equation}F^{ij}=(E^{-1})^{ij}=\delta_{ij} -\Pi^{ij}\end{equation}
It is easy to see that there is an interval of $a$'s for which the matrix $F^{ij}$ can be inverted.

Finally the action is

\begin{equation}S=\int dzd\bar z \bigl\{E_{ij}\partial x^i\bar\partial x^j +\alpha(x^2-1)\bigr\}\label{su2}\end{equation}
Notice that our model satisfies Eq.(\ref{sym}) for both, left and right invariant vector fields $v_a$.

\section{The dual model}
To write the dual model we need to describe the Poisson-Lie structure on solv(sl(2,c)$^r$). Since the bialgebra structure is not coboundary we can not use directly the Sklyanin bracket. Nevertheless, as remarked in section 3, the double, in this case sl(2,c)$^r$, has also a bialgebra structure which is always coboundary and whose $r$-matrix is given by

\begin{equation}
r={1\over 2}(G\wedge\tilde G+E\wedge\tilde E+F\wedge\tilde F)
\end{equation} 
The Poisson structure in the double is now given by the Sklyanin bracket. Since solv(sl(2,c)$^r$) is a Poisson-Lie soubgroup of sl(2,c)$^r$ (except for a global minus sign) the Poisson bivector on solv(sl(2,c)$^r$) is given by (minus) the restriction of the Poisson bivector in the double.

solv(sl(2,c)$^r$) is parametrized by exponential coordinates
\begin{equation}
h=e^{h^0\tilde G}e^{h^1\tilde E}e^{h^2\tilde F}=e^{-h^0\over 2}\pmatrix{ 1 & -h^2+ih^1\cr 0  & e^{h^0} }\end{equation}
The multiplication of the bialgebra structure by a constant $a$ as in Eq.(\ref{pbi}) amounts here to a dilatation of the generators of the dual group ($\tilde X\mapsto a\tilde X$). The coordinates $h^i$ are then the corresponding rescaled coordinates.
 In the basis $\{\partial_{h^0},\partial_{h^1},\partial_{h^2}\}$ the Poisson bivector is the matrix
\begin{equation}
\Pi^{ij}_h=-\pmatrix{0 & 2h^2 & 2h^1\cr -2h^2 &  0 &  ((h^1)^2+(h^2)^2+1)-e^{2h^0}\cr -2h^1 & -((h^1)^2+(h^2)^2+1)+e^{2h^0} & 0}\end{equation}

The other term in the action is precisely the inverse of the natural invariant metric on SU(2) given by the Killing form (we take $K_{ij}=-{\hbox{Tr}}X_iX_j/2=\delta_{ij}$ where $X_i$ are the SU(2) generators). We are dualizing with respect to the left invariant vector fields, that is, the generators of the right translations. The dual model is given by
\begin{equation}S=\int dzd\bar z \bigl\{(K^{-1} - \Pi^R_h)^{-1}_{ij}(\partial h h^{-1}\bar\partial hh^{-1}
\bigr\}\end{equation}
or, in terms of $\Pi_h$

\begin{equation}S=\int dzd\bar z \bigl\{(A_h - \Pi_h)^{-1}_{ij}\partial h^i\bar\partial h
^j )\bigr\}\end{equation}
where 
\begin{equation}
A_h=\pmatrix{1& 0 & 0\cr 0 & e^{2h_0} & 0 \cr 0 & 0 & e^{2h_0}}
\end{equation}

It is easy to see that the matrix $A_h - \Pi_h$ is invertible for all the range of values of $h^i$.

\begin{figure}
\begin{center}
~\epsfig{file=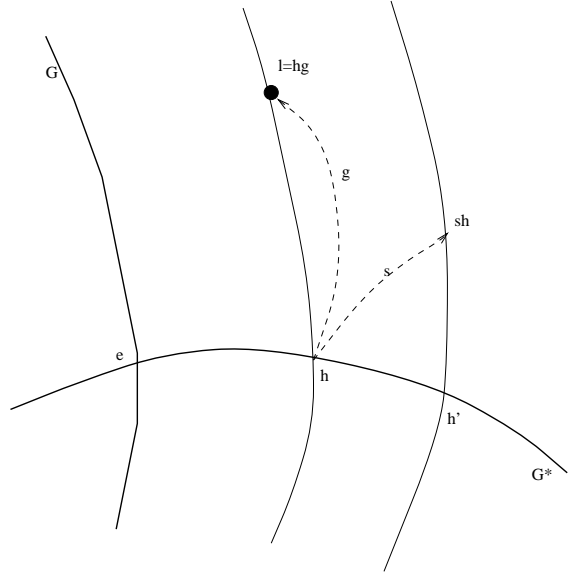,width=3in,angle=270}
\end{center}
\caption{Dressing transformations.}
\end{figure}

One way of relating both models is through the dressing transformations. An element $l$ of the double $D$ is uniquely decomposed as $l=\tilde g \tilde h$ where $\tilde g\in G$, $\tilde h\in G^*$. In our case, this decomposition is global since it corresponds to the Iwasawa decomposition on sl(2,c)$^r$, so $D=G\times G^*$ as manifolds. We have then on $D$ a structure of principal fibre bundle over $G$ with structure group $G^*$. The decomposition can also be written as $l= hg$, and in this case we are viewing $D$ as a principal fibre bundle over $G^*$ with structure group $G$. The right action of $G$ on $D$ moves us along the fiber. But we can define a left action of $G$ on $G^*$. To each point $h \in G^*$ and $s\in G$ we associate the projection $h'=\pi_{G^*}(sh)\in G^*$ (see Figure 1). It is easy to show that this is a well defined left action of $G$ on $G^*$. The orbits of this action are the symplectic leaves of the Poisson structure on $G^*$.
Let $\tilde g(z,\bar z)\in G$ be an extremal surface of the original model
(\ref{su2}), and let $\tilde h(z,\bar z)\in G^*$  be a solution of (\ref{bi}) for the same model, that is,
\begin{equation}
J(z,\bar z)=\tilde h(z,\bar z)d\tilde h^{-1}(z,\bar z)
\end{equation}
Following \cite{ks1} one can lift the extremal surface to $D=G\times G^*$  by $l(z,\bar z)=\tilde g(z,\bar z)\tilde h(z,\bar z)$. The projection of $l(z,\bar z)$ onto $G^*$ by means of the opposite decomposition $l(z,\bar z)=h(z,\bar z)g(z,\bar z)$ defines  an extremal surface $h(z,\bar z)$ of the dual model. So we see that "on shell" dressing transformations have the property of sending a solution to the conservation law in the original model to a extremal surface in the dual model. In this way, we realize the property of duality transformations, that is, they interchange Bianchi identities with equations of motion.

For completeness, we give here the dressing transformations of SU(2) acting on solv(sl(2,c)$^r$. We parametrize SU(2) by exponential coordinates
\begin{equation}
g(g^0,g^1,g^2)=e^{g^0G}e^{g^1E}e^{g^2F}
\end{equation}
Each of the uniparametric subgroups acts on $h\in$ solv(sl(2,c)$^r$ as
\subparagraph{1. $g=e^{tG}$} 
\begin{equation}
\matrix{\tilde h^0=h^0\cr
\tilde h^1= -h^2\sin(2t) +h^1\cos(2t)\cr
\tilde h^2= h^2\cos(2t) +h^1\sin(2t)}
\end{equation}
and  the infinitesimal transformation is given by
\begin{equation}
X_G=h^0\partial{h^0}+h^1\partial{h^2}-h^2\partial{h^1}
\end{equation}
\subparagraph{2. $g=e^{tE}$}
\begin{equation}
\matrix{ e^{\tilde h^0}=e^{-h^0}\bigl[\sin^2t\bigl(1+(h^1)^2+(h^2)^2\bigr)\bigr]+
e^{h^0}\cos^2t+2(\cos t\sin t) h^2\cr
\tilde h^1=h^1\cr
\tilde h^2= \cos t\sin t \bigl[e^{-h^0}\bigl(1+(h^1)^2+(h^2)^2\bigr)-e^{h^0}\bigr] +(\cos^2t-\sin^2t)h^2}
\end{equation}
and the infinitesimal transformation is
\begin{equation}
X_E=2e^{-h^0}h_2\partial_{h^0}+\bigl[e^{-h^0}\bigl(1+(h^1)^2+(h^2)^2\bigr)-e^{h^0}\bigr]\partial_{h^2}
\end{equation}
\subparagraph{3. $g=e^{tF}$}
\begin{equation}
\matrix{e^{\tilde h^0}=\sin^2te^{-h^0}\bigl(1+(h^1)^2+(h^2)^2\bigr) +e^{h^0}\cos^2t-2h^1\sin t\cos t\cr
\tilde h^1=-\cos t\sin t \bigl[e^{-h^0}\bigl(1+(h^1)^2+(h^2)^2\bigr)-e^{h^0}\bigr]+(\cos^2t-\sin^2t)h^1\cr
\tilde h^2==h^2}
\end{equation}
and the infinitesimal transformation
\begin{equation}
X_F=-2h^2e^{-h^0}\partial_{h^0}+\bigl[e^{-h^0}\bigl(1+(h^1)^2+(h^2)^2\bigr)-e^{h^0}\bigr]\partial_{h^2}
\end{equation}
\section{Conclusions}

In this letter we have explicitly constructed a dual pair of sigma models related by Poisson-Lie symmetry. The target manifolds are respectively SU(2) and SL(2,c)$^r/SU(2) \approx$SO(3,1)/SO(3). The models are complicated and no solutions to them are known, but since the framework of Poisson-Lie symmetry seems to be the appropriate one to define duality transformartions, we believe that non-trivial examples may be useful in future applications. In particular, the quantization of these models may give rise to a connection with SU(2)$_q$, problem that at this moment remains open. 

\section{Aknowledgments}
We want to thank A.C. Cadavid for useful discussions.

\section{Note}

After writting this letter we were informed of a preprint \cite{s} where a similar model was considered in the context of the renormalization group.

\end{document}